\pdfoutput=1

\documentclass[]{spie}  

\usepackage{amsmath,amsfonts,amssymb}
\usepackage{graphicx}
\usepackage{subfig}
\usepackage{enumitem}
\usepackage[colorlinks=true, allcolors=blue]{hyperref}

\title{Cooldown Strategies and Transient Thermal Simulations for the Simons Observatory}

\author[a]{Gabriele Coppi}
\author[a]{Zhilei Xu}
\author[j]{Aamir Ali}
\author[a]{Mark J. Devlin}
\author[a]{Simon Dicker}
\author[d]{Nicholas Galitzki}
\author[e]{Patricio A. Gallardo}
\author[d]{Brian Keating}
\author[a]{Michele Limon}
\author[g]{Marius Longu}
\author[c]{Andrew J. May}
\author[b]{Jeff McMahon}
\author[e]{Michael D. Niemack}
\author[a]{Jack L. Orlowski-Scherer}
\author[c]{Lucio Piccirillo}
\author[f]{Giuseppe Puglisi}
\author[k]{Maria Salatino}
\author[b]{Sara M. Simon}
\author[d]{Grant Teply}
\author[a,h]{Robert Thornton}
\author[e]{Eve M. Vavagiakis}
\author[a]{Ningfeng Zhu}
\affil[a]{Department of Physics \& Astronomy, University of Pennsylvania, Philadelphia, Pennsylvania, PA, USA}
\affil[b]{Department of Physics, University of Michigan, Ann Arbor, USA}
\affil[c]{Jodrell Bank Centre for Astrophysics, University of Manchester, Manchester, UK}
\affil[d]{Department of Physics, UCSD, La Jolla, CA, USA}
\affil[e]{Department of Physics, Cornell University, Ithaca, NY USA}
\affil[f]{Department of Physics, Stanford University, Stanford, California, CA}
\affil[g]{Department of Physics, Princeton University, Princeton, NJ, USA}
\affil[h]{Department of Physics \& Engineering, West Chester University of Pennsylvania, West Chester, PA, USA}
\affil[j]{Department of Physics, University of California, Berkeley, Berkeley, CA, USA}
\affil[k]{AstroParticle and Cosmology (APC) laboratory, Paris Diderot University, Paris, France}

\authorinfo{Further author information: (Send correspondence to Gabriele Coppi)\\Gabriele Coppi: E-mail: gcoppi@sas.upenn.edu}
\begin{document} 
\maketitle

\begin{abstract}
The Simons Observatory (SO) will provide precision polarimetry of the cosmic microwave background (CMB) using a series of telescopes which will cover angular scales from arc-minutes to tens of degrees, contain over 60,000 detectors, and observe in frequency bands between 27 GHz and 270 GHz. SO will consist of a six-meter-aperture telescope initially coupled to roughly 35,000 detectors along with an array of half-meter aperture refractive cameras, coupled to an additional 30,000+ detectors. 

The large aperture telescope receiver (LATR) is coupled to the SO six-meter crossed Dragone telescope and will be 2.4 m in diameter, weigh over 3 metric tons, and have five cryogenic stages (80 K, 40 K, 4 K, 1 K and 100 mK). The LATR is coupled to the telescope via 13 independent optics tubes containing cryogenic optical elements and detectors. The cryostat will be cooled by by two Cryomech PT90 (80 K) and three Cryomech PT420 (40 K and 4 K) pulse tube cryocoolers, with cooling of the 1 K and 100 mK stages by a commercial dilution refrigerator system. The secondo component, the small aperture telescope (SAT), is a single optics tube refractive cameras of 42 cm diameter. Cooling of the SAT stages will be provided by two Cryomech PT420, one of which is dedicated to the dilution refrigeration system which will cool the focal plane to 100 mK. SO will deploy a total of three SATs.

In order to estimate the cool down time of the camera systems given their size and complexity, a finite difference code based on an implicit solver has been written to simulate the transient thermal behavior of both cryostats. The result from the simulations presented here predict a 35 day cool down for the LATR. The simulations suggest additional heat switches between stages would be effective in distribution cool down power and reducing the time it takes for the LATR to reach its base temperatures. The SAT is predicted to cool down in one week, which meets the SO design goals.

\end{abstract}


\section{INTRODUCTION}\label{sec:intro}

The cosmic microwave background (CMB) has become one of the most powerful probes of the early universe. Measurements of its temperature anisotropy on the level of $\sim$ ten parts per million, which have brought cosmology into a precision era, have placed tight constraints on the fundamental properties of the universe. Beyond the temperature anisotropy, CMB polarization anisotropy not only enriches our understanding of our cosmological model, but could potentially provide clues to the very beginning of the universe via the detection (or non-detection) of primordial gravitational waves.  A number of experiments have made and are continuing to refine measurements of the polarization anisotropy.  However, these experiments are typically dedicated to a relatively restricted range of angular scales, e.g., large angular scales (tens of degrees) or high resolution/small angular scales ($\sim$~1 arcminute). To provide a complete picture of cosmology, both large and small angular scales are important. Ideally these measurements would be made from the same observing site so that the widest range of angular scales can be probed, at multiple frequencies, on the same regions of the sky.  This is the goal of the Simons Observatory (SO).  

The Simons Observatory will comprise a combination a single large aperture telescope receiver (LATR) and and array of small aperture telescopes (SAT) for observing both large and small angular scales. The observatory will be located in Chile's Atacama Desert at an altitude of 5190 m. The LATR is designed with a large FOV capable of supporting a cryostat with up to 19 optics tubes.  To limit the development risk, the LATR is designed to accommodate up to 13 optics tubes.  We plan to deploy 7 optics tubes with 3 detector wafers in each for a total of roughly 35,000 detectors, primarily at 90/150 GHz in the initial SO deployment.  We note that each optics tube could be upgraded to support 4.5 wafers for a~50\% increase in the number of detectors per optics tube. With this upgrade and the deployment of 19 optics tubes, the LAT could support roughly 145,000 detectors at 90/150 GHz. In addition, the SAT cameras will be deployed with more than 30,000 detectors. With the current scale and potential to upgrade, SO will serve as a valuable step to advance to the next-generation CMB experiment, CMB-S4 \citenum{Technology, Science}.

To achieve the scientific goal of Simons Observatory, it is important to reduce the non observing time and in particular the time required for cooling down the cryostat as part of the preparation. In this paper, we discuss a model developed to estimate the cooldown of both the LATR and the SAT. In sections~\ref{sec:theory}\textendash\ref{sec:contacts}, the theory for the model is introduced and its mathematical implementation is explained. Sections \ref{sec:cooling} and \ref{sec:setup} are dedicated to introduce the general setup  of the system that we simulated including the cooling capacity available and the external load. Finally, in section \ref{sec:results} we discuss the results from the simulations. The model developed in this paper can be also useful in developing the future large CMB experiments, like CMB-S4.

\section{THEORY}\label{sec:theory}

To create a transient thermal model, it is necessary to introduce the basic equation for heat transfer. The 3D heat transfer equation in cartesian coordinates is presented in eq. \ref{eq:cartHT} while eq. \ref{eq:cylHT} shows the heat transfer equation in the cylindrical coordinates. In this equation $C$ is the heat capacity, $\rho$ is the density and $k$ is the thermal conductivity. $I$ is the internal heat generation expressed in W/m$^3$.

\begin{equation}
    \label{eq:cartHT}
    C\rho\frac{\partial T}{\partial t}=\frac{\partial}{\partial x}\left(k \frac{\partial T}{\partial x}\right)+\frac{\partial}{\partial y}\left(k \frac{\partial T}{\partial y}\right)+\frac{\partial}{\partial z}\left(k \frac{\partial T}{\partial z}\right) + I
\end{equation}

\begin{equation}
    \label{eq:cylHT}
    C\rho\frac{\partial T}{\partial t}=\frac{\partial}{r\partial r}\left(kr \frac{\partial T}{\partial r}\right)+\frac{\partial}{r^2 \partial \phi}\left(k \frac{\partial T}{\partial \phi}\right)+\frac{\partial}{\partial z}\left(k \frac{\partial T}{\partial z}\right) + I
\end{equation}

In general, these two equations do not have an explicit solution unless for particular cases. Therefore, to solve these equations it is necessary to discretize them as explained in the next section.

\section{DISCRETIZATION}\label{sec:discretization}

An implicit scheme based on \emph{finite-difference model} (FDM) implementation has been chosen to solve the heat transfer equations introduced in section \ref{sec:theory}. Using an implicit scheme as opposed to an explicit one reduces the instability of the solution. Because an implicit scheme considers the system both at the current and the final status while solving the equation. Meanwhile, the explicit scheme considers only the current status. In order to optimize the discretization of the problem, the heat transfer equation has been approached differently for each of the geometries considered. The primary geometries present in the physical camera models we consider can be divided into three categories:

\begin{itemize}
    \item $1$D rod, \\
    \item $2$D plate with a thickness significantly smaller than the radius, \\
    \item $2D$ cylinder whose thickness is significantly smaller than the radius.
\end{itemize}

The three geometries considered are representative of the current model of the large aperture receiver telescope described in \citenum{Zhu2018, Galitzki2018}.

\subsection{$1$D Rod}\label{subsec:1Drod}

The 1D rod geometry is used to describe heat straps. The heat straps for SO will be made from OFHC (oxygen-free high conductivity) copper with a high thermal conductivity which allows us to neglect the radial and azimuthal heat flows In this way, we only consider conduction along the z-axis, or along the main axis of the cylinder. 

With a single conduction axis eq. \ref{eq:cylHT} becomes one dimensional. According to the implicit FDM that has been adopted, the heat transfer equation can be discretized as follows:

\begin{equation}
    \label{eq:1D_FDM}
    C^{n}_i\rho_i\frac{T_{i}^{n+1}-T_{i}^{n}}{\Delta t}=\frac{k_{i+1/2}^n\left(T^{n+1}_{i+1}-T^{n+1}_{i}\right)-k_{i-1/2}^n\left(T^{n+1}_{i}-T^{n+1}_{i-1}\right)}{\left(\Delta z\right)^2}
\end{equation}

where:

\begin{subequations}
\begin{equation}
  k_{i+1/2} = \frac{k_{i+1}+k_{i}}{2} 
\end{equation}    
\begin{equation}
  k_{i-1/2} = \frac{k_{i-1}+k_{i}}{2} 
\end{equation}
\label{eq:k}
\end{subequations}

The N equations resulting from the discretization form a system that can be written as:

\begin{equation}
    \label{eq:matrix}
    A T^{n+1} = T^n
\end{equation}

where A is the matrix of the coefficients and can be written as:

\begin{equation}
    \label{eq:matrix1D}
    A = 
    \begin{pmatrix}
     \cdots & \cdots & \cdots & \cdots  & \cdots & \cdots\\
     A_{2,1} & B_{2,2} & C_{2,3} & \cdots & \cdots & 0\\
     0 & A_{3,2} & B_{3,3} & C_{3,4} & \cdots & 0 \\
     \vdots  & \vdots  &  \vdots & \ddots & \vdots  & \vdots\\
     \cdots & \cdots & \cdots & \cdots & \cdots & \cdots
 \end{pmatrix}
\end{equation}

where:

\begin{subequations}
\begin{equation}
  A_{i,j} = -\frac{\Delta t}{2C^{n}_i\rho_i \left(\Delta z\right)^2}\left(k_{i-1}+k_{i}\right)
\end{equation}    
\begin{equation}
  B_{i,j} = 1+\frac{\Delta t}{2C^{n}_i\rho_i \left(\Delta z\right)^2}\left(k_{i+1}+2k_{i}+k_{i-1}\right)
\end{equation}
\begin{equation}
  C_{i,j} = -\frac{\Delta t}{2C^{n}_i\rho_i \left(\Delta z\right)^2}\left(k_{i+1}+k_{i}\right)
\end{equation}  
\label{eq:matrix1Dcoefficients}
\end{subequations}

In the matrix in eq. \ref{eq:matrix1D}, the first and the last rows are empty because they need to be filled with the parameters coming from the boundary conditions. If the boundary condition is a Dirichlet condition\footnote{This boundary condition means that the temperature on the boundary is constant.}, the row is filled with a $0$ except the first (or the last) element which is $1$. Instead for a Neumann boundary condition\footnote{This boundary condition means that the first derivative of the temperature, so the heat flow, on the boundary is constant.} it is possible to write the energy balance at these nodes as:   

\begin{subequations}
\begin{equation}
  mC(T)\frac{dT}{dt} = \dot{Q}^n+\frac{A_r}{\Delta z}\int_1^2 k(T)dT
\end{equation}    
\begin{equation}
  mC(T)\frac{dT}{dt} = \dot{Q}^n+\frac{A_r}{\Delta z}\int_{m-1}^m k(T)dT
\end{equation}
\label{eq:1DBC}
\end{subequations}

where $A_r$ is the area of the rod and $m$ is the mass of the element of length $\Delta z$. The eqns. \ref{eq:1DBC} can be discretized using a Taylor approximation and the missing matrix elements become:

\begin{subequations}
\begin{align}
    B_{1,1} &= 1+\frac{\Delta t}{2C^{n}_i\rho_i \left(\Delta z\right)^2}\left(k_{1}+k_{2}\right) \\
    C_{1,2} &= -\frac{\Delta t}{2C^{n}_i\rho_i \left(\Delta z\right)^2}\left(k_{1}+k_{2}\right) \\
    B_{m,m} &= 1+\frac{\Delta t}{2C^{n}_i\rho_i \left(\Delta z\right)^2}\left(k_{m}+k_{m-1}\right)\\
    A_{m,m-1}&=-\frac{\Delta t}{2C^{n}_i\rho_i \left(\Delta z\right)^2}\left(k_{m}+k_{m-1}\right)
\end{align}
\label{eq:BC-FDM1Dcoeff}
\end{subequations}

However, the additional $\dot{Q}$ term in eqns. \ref{eq:1DBC} modifies eq. \ref{eq:matrix} which  becomes:

\begin{equation}
    \label{eq:matrix2}
    AT^{n+1}=T^n+l
\end{equation}

where $l$ is a vector of all zeros except for the first and last elements which are:

\begin{subequations}
    \begin{equation}
        \label{eq:load_road1}
        l[1] = -\frac{\dot{Q}}{\Delta z A_r C_1 \rho_1}
    \end{equation}
    \begin{equation}
        \label{eq:load_road2}
        l[m] = -\frac{\dot{Q}}{\Delta z A_r C_m \rho_m}
    \end{equation}
    \label{eq:ln}
\end{subequations}

The heat flow $\dot{Q}$ can be dependent on the temperature of the element. For example, the cooling power of every mechanical cooler is dependent on the temperature of the element which is in contact with it. In this case,  this needs to be computed at the instant $n+1$ and the resulting element needs to move on the left side of the system \ref{eq:matrix2}.

\subsection{2D Plate}\label{subsec:plate}

A 2D plate geometry is used to compute the heat transfer across plates, like filter plates or cylinder caps. Since the radius is significantly larger than the thickness, the plate is considered to be isothermal along the z-axis. With this simplification, the heat transfer equation \ref{eq:cartHT} becomes:

\begin{equation}
    \label{eq:2DHT-plate}
    C\rho\frac{\partial T}{\partial t}=k\frac{\partial}{\partial x}\left(\frac{\partial T}{\partial x}\right)+k\frac{\partial}{\partial y}\left(\frac{\partial T}{\partial y}\right) + I
\end{equation}

For this geometry, the internal heat source is not assumed to be zero as an external load can be along the z axis, e.g. on the face of the plane. In this scenario we consider the external load as an internal heat source. This is possible as the conservation of energy implies an internal heat source equivalent to an external heat source will warm or cool the element it is coupled to by the same amount.

Similarly to the 1D rod model, the 2D plate geometry can be discretized using a finite difference method, using an implicit scheme. In this case, eq. \ref{eq:2DHT-plate} becomes:

\begin{equation}
    \label{eq:2D_plate_FDM}
    \begin{split}
    & T_{i,j}^{n+1}\left(1+\frac{\Delta t }{C^{n}_{i,j}\rho_{i,j}}\left(\frac{k_{i+1/2,j}^n F_{i+1,j}+k_{i-1/2,j}^n F_{i-1,j}}{\left(\Delta x\right)^2}+\frac{k_{i,j-1/2}^n F_{i,j-1}+k_{i,j+1/2}^n F_{i,j+1}}{\left(\Delta y\right)^2}\right)\right)-\\
    &T_{i+1,j}^{n+1}\left(\frac{\Delta t k_{i+1/2,j}^n F_{i+1,j}}{C^{n}_{i,j}\rho_{i,j} \left(\Delta x\right)^2}\right)- T_{i-1,j}^{n+1}\left(\frac{\Delta t k_{i-1/2,j}^n F_{i-1,j}}{C^{n}_{i,j}\rho_{i,j} \left(\Delta x\right)^2}\right)-\\
    &T_{i,j+1}^{n+1}\left(\frac{\Delta t k_{i,j+1/2}^n F_{i,j+1}}{C^{n}_{i,j}\rho_{i,j} \left(\Delta y\right)^2}\right)- T_{i,j-1}^{n+1}\left(\frac{\Delta t k_{i,j-1/2}^n F_{i,j-1}}{C^{n}_{i,j}\rho{i,j}i \left(\Delta y\right)^2}\right) = \\
    & T^{n}_{i,j}+\frac{Q_{i,j}}{C^{n}_{i,j}\rho_{i,j}}
    \end{split}
\end{equation}

where $F$ is a function that is equal to $1$ if the cell is part of the plate and $0$ if the cell is not. This function automatically enforces a Neumann boundary condition on each edge, fixing the heat flow at $0$.
The thermal conductivity $k$ is computed according to eq. \ref{eq:k}.  
As in the previous case, $k$ and $C$ are computed at the instant $n$.

Using a cartesian geometry instead of a polar geometry made it easier to define the positions of the filters and optics tubes in the model. 
Similarly to the 1D rod case, it is possible to write a coefficient matrix which is equal to:

\begin{equation}
    \label{eq:matrix2D_plate}
    A = 
    \begin{pmatrix}
     1 & 0 & \cdots & \cdots & \cdots & \cdots & \cdots & \cdots & \cdots  & 0\\
     0 & 1 & \cdots & \cdots & \cdots & \cdots & \cdots & \cdots & \cdots  & 0\\
     \vdots  & \vdots & \ddots  & \vdots & \vdots & \vdots  & \vdots  & \vdots & \vdots  & \vdots\\
     \cdots & \cdots & C_{y, y-x} & \cdots & A_{y,y-1} & B_{y,y} & A_{y,y+1} & \cdots & C_{y, y+x} &\cdots  \\
     \vdots  & \vdots & \vdots  & \vdots & \vdots & \vdots  & \vdots  & \vdots & \ddots  & \vdots\\
      0 & \cdots & \cdots & \cdots & \cdots & \cdots & \cdots & \cdots & \cdots & 1\\
 \end{pmatrix}
\end{equation}

where $x$ is the number of elements along the x-axis and $y$ is an index to convert the 2D plate array into a 1D array, $y=x*i+j$. The matrix coefficients are defined as:

\begin{subequations}
\begin{align}
    B_{y,y} &\begin{aligned}[t]
    =& 1+\frac{\Delta t }{C^{n}_{i,j}\rho_{i,j}}\left(\frac{k_{i+1/2,j}^nF_{i+1,j}+k_{i-1/2,j}^nF_{i-1,j}}{\left(\Delta x\right)^2}\right.\\
    &\left. +\frac{k_{i,j-1/2}^nF_{i,j-1}+k_{i,j+1/2}^nF_{i,j+1}}{\left(\Delta y\right)^2}\right) \\
    \end{aligned} \\
    C_{y,y\pm x} &= -\frac{\Delta t k_{i,j\pm 1/2}^nF_{i\pm 1,j}}{C^{n}_{i,j}\rho_{i,j} \left(\Delta y\right)^2} \\
    A_{y,y\pm1} &= \frac{\Delta t k_{i\pm 1/2,j}^nF_{i\pm 1,j}}{C^{n}_{i,j}\rho_{i,j} \left(\Delta x\right)^2}
\end{align}
\label{eq:BC-FDM2Dcoeff_plate}
\end{subequations}

The matrix equation that needs to be solved is the following:

\begin{equation}
    \label{eq:2Dmat_plate}
    (A+B)T^{n+1} = T^{n}
\end{equation}

where $B$ is a matrix whose elements is given by $\frac{I_{i,j}^{n+1}}{C_{i,j}\rho_{i,j}}$.

In this case, $I$ has a linear dependency on $T$ which is why it is computed at the instant $n+1$.

\subsection{2D Cylinder}\label{subsec:cyl}

The 2D cylinder geometry is used to compute the heat transfer across thin walled cylindrical shells. Since the radius is significantly larger than the thickness, the cylinder is considered to be isothermal radially. 
This case is the same to the one described in subsection \ref{subsec:plate}, with the only difference that the equation that needs to be discretized is eq. \ref{eq:cylHT} without the partial derivative of the $r$ component. For this reason, the cylindrical discretization is not described in the same depth as the previous geometries.

\section{CONTACTS BETWEEN COMPONENTS}\label{sec:contacts}

Each stage of the cryostat will not be manufactured as a monolithic stage and as a consequence the stages will be composed of multiple components with contact interfaces. Each component can be modeled using one of the geometries described in the previous section. The interface between two bodies introduces an impedance to heat flow, which is called \emph{thermal contact resistance}. 

\subsection{Contact Theory}\label{subsec:contheory}

Estimating the thermal contact resistance is complex as it involves many factors such as the contact pressure, the smoothness of the surface in contact and the materials of the components. As shown in \citenum{Salerno1997}, some measurements of contact resistance have been made at cryogenic temperatures. However, these measurements do not include common combinations of contact materials, e.g. aluminum and copper. Moreover, these measurements are valid only at low temperatures, whereas for a cooldown estimation it is necessary to know the value from room temperature. 

In order to estimate the thermal conductance, defined as the reciprocal of the thermal resistance, it is possible to use the \emph{Cooper-Mikic-Yovanovich} correlation, ref. \citenum{Yovanovich2003}. According to this model, the conductance can be computed as: 

\begin{equation}
    \label{eq:TCRtheory}
    h_c = 1.25\frac{Akm}{\sigma}\left(P/H_c\right)^{0.95}
\end{equation}

where $A$ is the contact area, $P$ is the contact pressure, $k$ is the harmonic mean,  $\sigma$ is the effective rms roughness and $m$ is the effective absolute mean asperity slope. The last three quantities are defined as:

\begin{equation}
    \label{eq:kmean}
    k = \frac{2k_1k_2}{k_1+k_2}
\end{equation}

\begin{equation}
    \label{eq:sigma}
    \sigma = \sqrt{\sigma_1^2+\sigma_2^2}
\end{equation}

\begin{equation}
    \label{eq:m}
    m = \sqrt{m_1^2+m_2^2}
\end{equation}

In eqns. \ref{eq:kmean},\ref{eq:sigma} and \ref{eq:m}, the subscripts $1$ and $2$ refer to the first and the second body in contact, respectively. Finally, the $H_c$ in the eq.\ref{eq:TCRtheory} is the lowest Vickers microhardness value of the materials in contact. 

The values of the Vickers microhardness is not available for each material. A theoretical formula for estimating the Vickers microhardness is given by:

\begin{equation}
    \label{eq:vickhard}
    H_c = P^{0.071^{c_2}}\left(\frac{1.62c_1(\sigma/m)^{c_2}}{P}\right)^{(1/(1+0.071^{c_2}))}
\end{equation}

where $c_1$ and $c_2$ are the Vickers coefficients for each material. These values can be found experimentally or by using the relationship that they have with the Brinell hardness. These relations are  expressed in eq. \ref{eq:brincoeff}, which gives the value of $c_1$ in MPa, while $c_2$ is dimensionsless.

\begin{subequations}
\begin{align}
  c_1 & = 3178\left(4.0 - 5.77H_B+4.0 H_B^2 - 0.61 H_B^3\right) \\
  c_2 &=-0.370+0.442 \frac{H_B}{c_1}
\end{align}
\label{eq:brincoeff}
\end{subequations}

\subsection{Implementation in the model}\label{subsec:implement}

The thermal contact conductance defined in eq. \ref{eq:TCRtheory} is important to compute the heat flow between two different component in contact. Indeed, the flow between two different components is given by:

\begin{equation}
    \label{eq:heatCont}
    \dot{Q} = h_c\left(T_1-T_2\right)
\end{equation}

where $h_c$ is the thermal conductance as defined in subsection \ref{subsec:implement}.

In order to maintain a simple and coherent mesh\footnote{In this case, mesh refers to the division of the components in multiple elements.} interface between multiple components, the contact has been considered localized in a single mesh cell for each component. 
Therefore, the heat flow between the plate and the heat strap is given by:

\begin{equation}
    \label{eq:thermcont_SP_1}
    \dot{Q} = h_c\left(T^{n+1}_s-T^{n+1}_p\right)
\end{equation}

from the plate point of view. Alternatively, from the strap, the flow is equal to:

\begin{equation}
    \label{eq:thermcont_SP_2}
    \dot{Q} = h_c\left(T^{n+1}_p-T^{n+1}_s\right)
\end{equation}.

In this case it is possible to notice the the temperatures $T$ are computed at the instant $n+1$. This is necessary since the heat flow is a function of temperature as explained in the previous section.

Eq. \ref{eq:thermcont_SP_1} is included in the $B$ matrix in eq. \ref{eq:2Dmat_plate} where the contact happens, replacing the internal heat generation. Whereas, eq.\ref{eq:thermcont_SP_2} is used at the end (or at the beginning, depending where the contact happens) of the vector $l$ of the eq.\ref{eq:matrix2}.  The resulting equations for the strap and the plate need to be combined together in order to be solved.

\subsection{Thermal contact between stages}\label{subsec:contactstages}

Two different stages in the cryostat can be thermally connected using a heat switch. This component help in transferring some cooling power from the higher cooling capacity stage to a lower one. 
The heat flow through a heat switch is given by:

\begin{equation}
    \label{eq:heatswitch}
    \dot{Q} = C\left(T_1-T_2\right)
\end{equation}

where $C$ is the conductance of the switch. Since  eq. \ref{eq:heatswitch} has the same form as eq. \ref{eq:heatCont}, the procedure expressed previously for the contact between component in a single stage can be applied here to connect different stages.

\section{COOLING POWER AND EXTERNAL LOAD}\label{sec:cooling}

\begin{figure}
\centering
  \subfloat[][First Stage]{\label{fig:PT420_first}\includegraphics[width=0.45\textwidth]{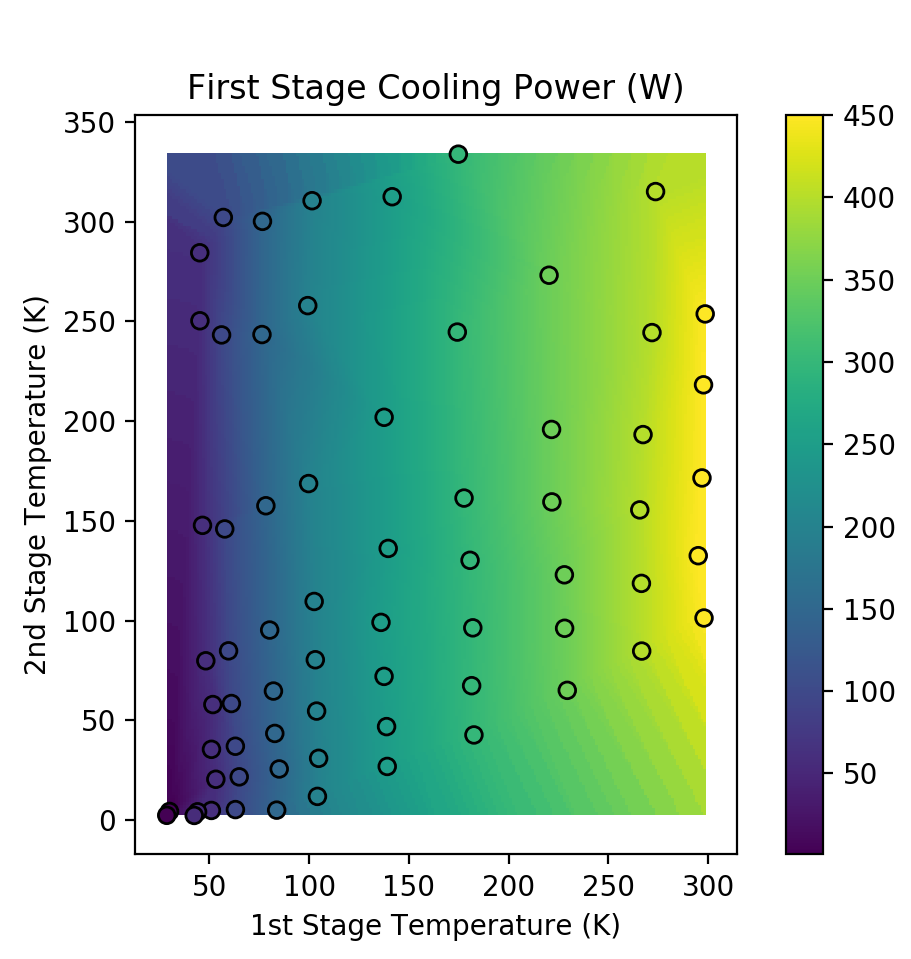}} \quad
    \subfloat[][Second Stage]{\label{fig:PT420_second}\includegraphics[width=0.45\textwidth]{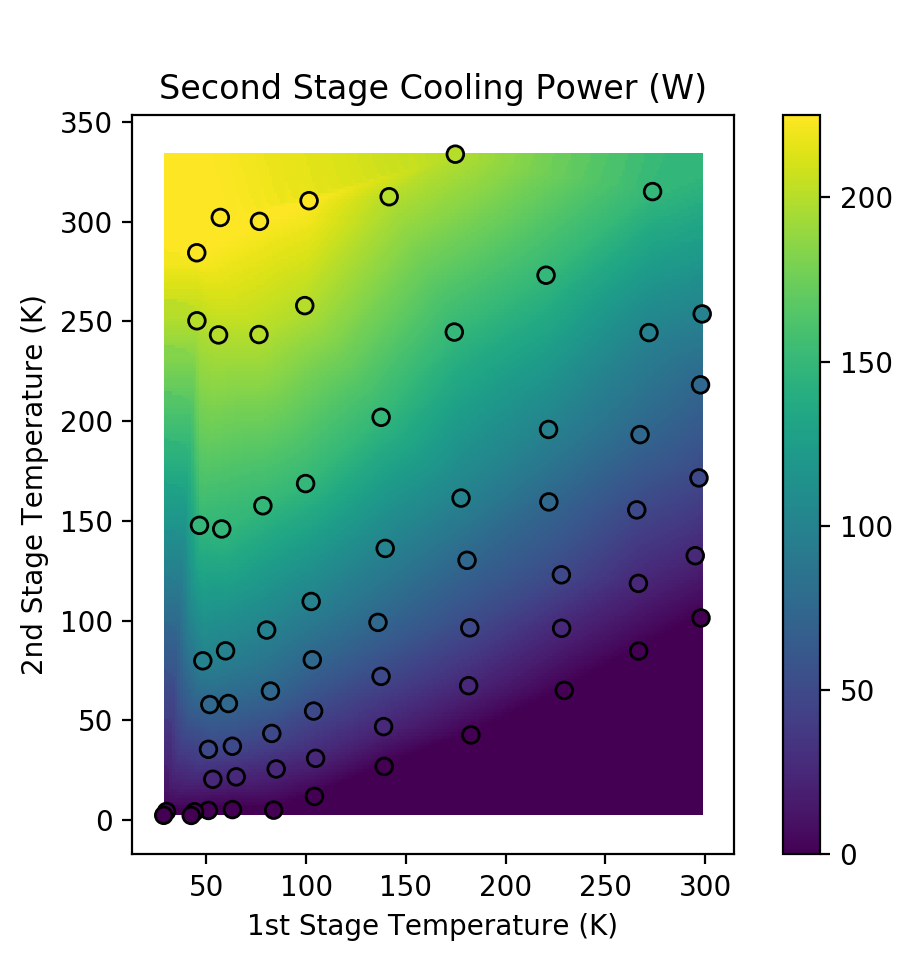}} \\
    \caption{Cooling power of the two stages of a PT420, ref. \citenum{CryomechPT420}. The dots are the value measured by Cryomech while the contour is a linear interpolation between these points. The color of the dots is proportional to the cooling power according to the same scale of the contour. Since the interpolation in the measured points differs only by a $10^{-3}$, the color of the dots is almost the same of the interpolation. From \citenum{CryomechPT420}.}
    \label{fig:PT420}
\end{figure}

\begin{figure}
    \centering
    \includegraphics[scale=0.6]{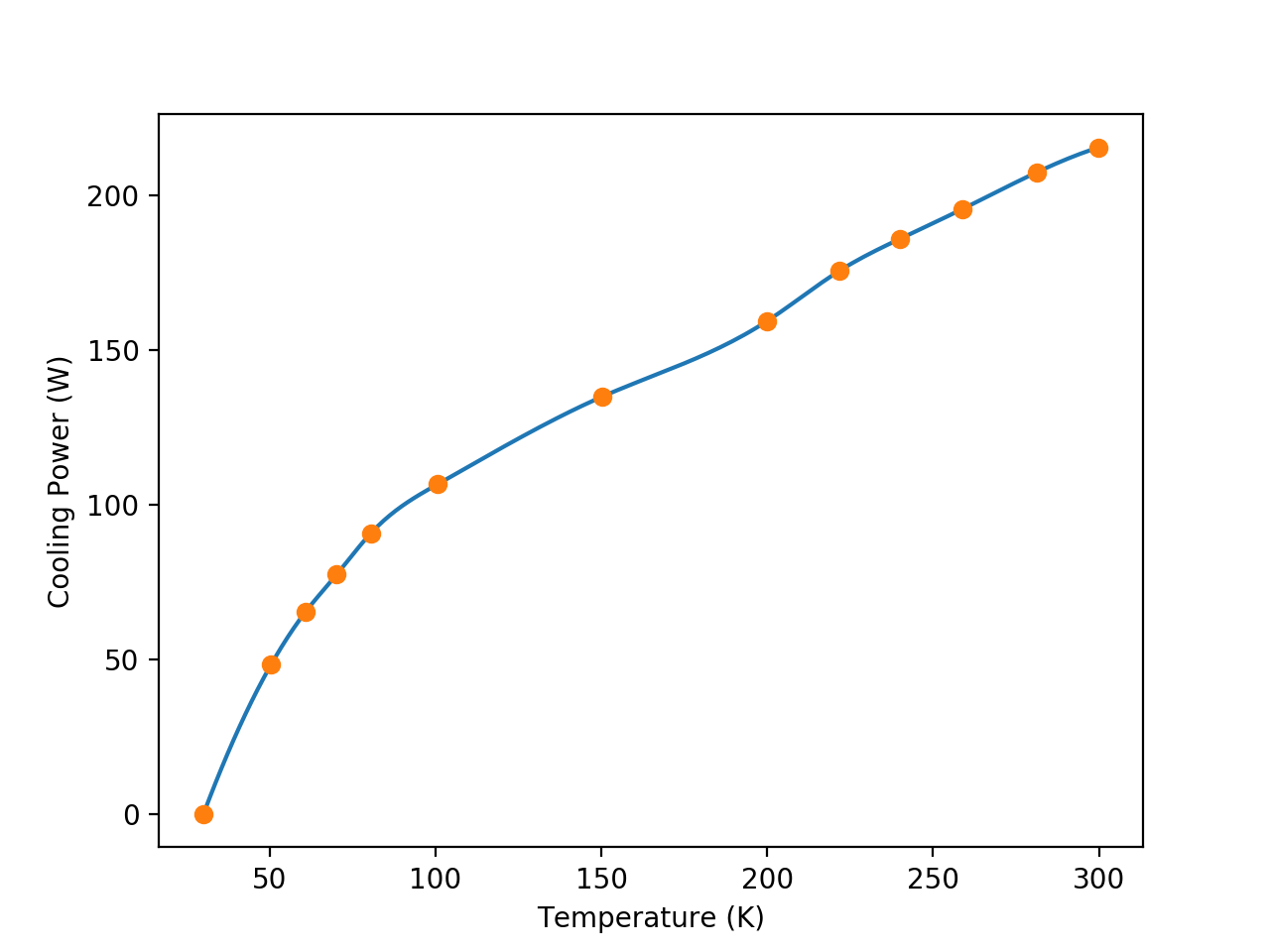}
    \caption{Cooling Power of the single stage PT90, ref. \citenum{CryomechPT420}. The dots are the value measured by Cryomech, while the line is the cubic interpolation used in the simulation code.}
    \label{fig:PT90}
\end{figure}

The cooling power on the different stages is provided by a series of mechanical coolers. In particular, the LATR uses Cryomech PT90s for the 80 K stage, Cryomech PT420s for the 40K and 4 K stages, and a Cryomech PT420 for the dilution refrigerator system. 
The cooling power for the PT420 the PT90 is presented in Fig. \ref{fig:PT420} and Fig \ref{fig:PT90}, respectively. 

The number of mechanical coolers for each stage has been computed considering the cooling power required during the operation. The static thermal model for the LATR is presented in \citenum{Orlowski-Scherer2018}. This model computes the load from different contributions on each stage. The main contributions to the load are: radiation from hotter stages, conduction through the support and the wires (for readout or housekeeping), filters, and loading produced by the cold readout electronics. The static thermal model is computed at different timesteps during the simulation to estimate the temporary load on each stage.

\begin{figure}[t]
    \centering
    \includegraphics[scale=0.4]{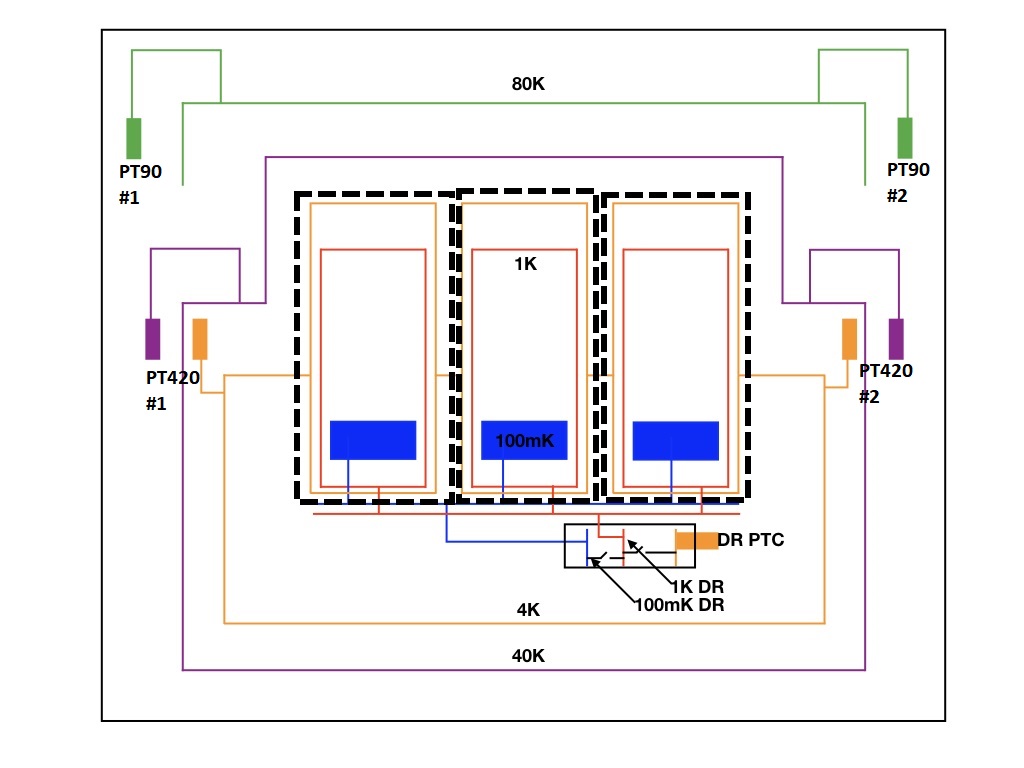}
    \caption{Schematic of the LATR cryostat model used in the simulation. The dashed lines enclose the optics tubes which have three different stages: the $4$ K, the $1$ K and finally the focal plane array which is represented by the blue box. The different mechanical cooler are labeled and the lines from the box representing the mechanical cooler to the stages are the heat straps. The path of the heat strap here is only representative to show a connection between the mechanical cooler and the stages, in reality the path will be different.}
    \label{fig:LATRmodel}
\end{figure}

\section{SIMULATION SETUP}\label{sec:setup}
The theory introduced in the previous sections was used to create a python code that can estimate the cooldown time for the LATR and the SAT. The code takes as input multiple parameters, including the geometry of each stage, the material of each component, the quantity and model of mechanical coolers, number of heat switches and their conductance and number of optics tube and magnetic shields. Moreover, each component's location in 3D space is defined through an additional input configuration file. 
For each time step the code first creates the matrix of each stage, then combines them in case of a presence of heat switches, and finally computes the temperature after a time $dt$ solving the generic matrix equation $AT^{n+1}=T^{n}$. The matrix $A$ introduced in section \ref{sec:discrtization} for each of the geometries considered is generally a very sparse matrix which allows us to utilize a sparse solver. Sparse solvers can handle large matrices and operate on them in a relatively short time. The code stops as soon as the temperature difference between the instants $n$ and $n+1$ for the elements is smaller than $10^{-3}$ K. This cutoff has been chosen because the error on the diodes thermometers down to $4$ K is on the same order of magnitude, therefore a simulation more accurate could not be verified in reality.

\subsection{Large Aperture Telescope Receiver}\label{subsec:LATRsetup}
For the LATR, the schematic of the geometry of the cryostat simulated is presented in Fig. \ref{fig:LATRmodel}. The heat straps between the mechanical coolers and the stages are made of OFHC copper for all of the stages. The material choice for each stage in the simulation are as follows: Al-6063 for the 80K stage, Al-6061 for the 40K, 4K stages and 1K optics tubes, OFHC copper for the 100mK stage and for the 1K thermal bus and finally Amumetal\footnote{http://www.amuneal.com/}, a nickel alloy, for the magnetic shields. The properties of the materials are taken from the NIST database\cite{nist2016}.
The cooling power is provided by two PT90 and two PT420 plus an additional PT420 dedicated specifically to the dilution refrigerator system.
For this simulation, we chose not to include any heat switches to compute the upper limit in the cool down time. Moreover, many of the commonly used heat switches do not have any conductivity data between room temperature and cryogenic temperatures which limits our ability to use them in cool down time simulations..

\subsection{Small Aperture Telescope}\label{subsec:SACsetup}
\begin{figure}[t]
    \centering
    \includegraphics[scale=0.35]{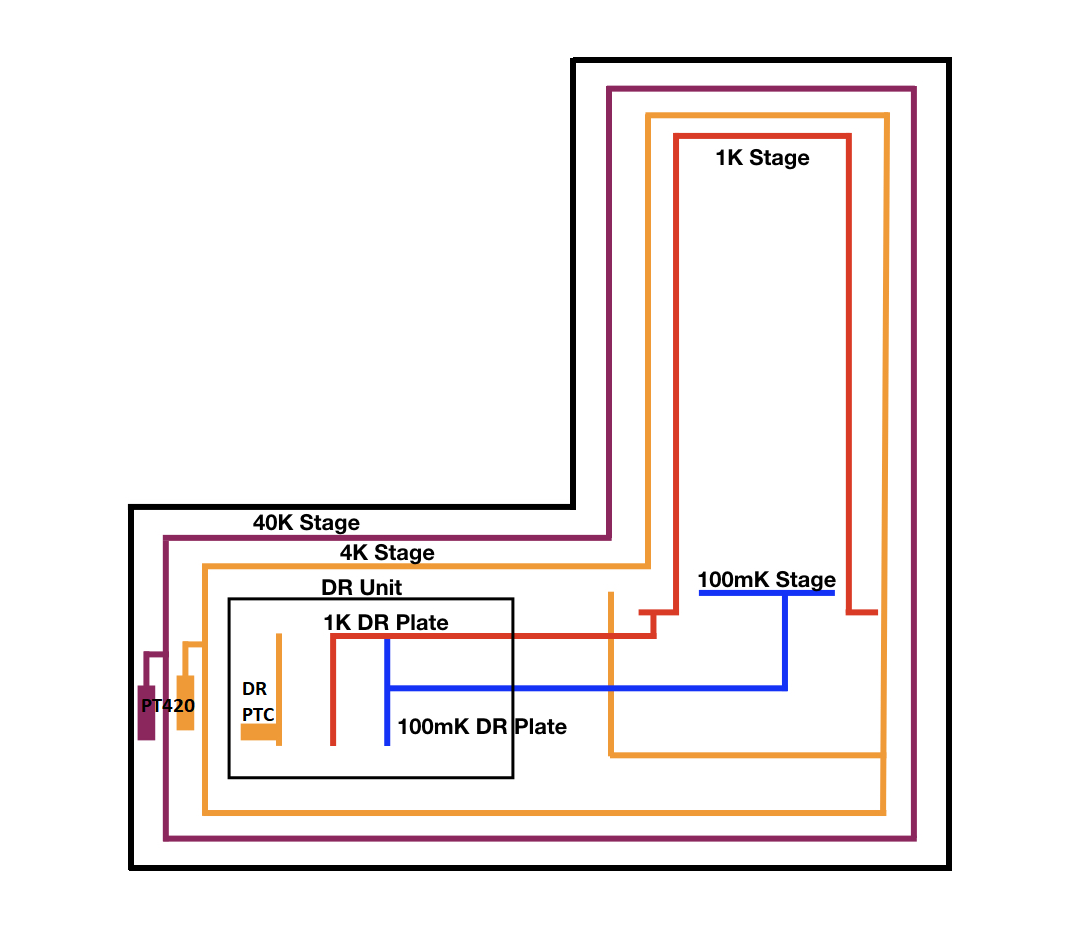}
    \caption{Schematic of the SAT cryostat model used in the simulation. The single optics tube is clearly visible on the right side of the figure. The cryogenics equipments is concentrated on the right side.}
    \label{fig:SMALLmodel}
\end{figure}

The SAT is composed of a single optics tube and no 80 K stage as is shown in the simulation schematic in Fig. \ref{fig:SMALLmodel}. Similarly to the LATR,  the 40K, 4K and 1K stages are simulated with Al-6061 set as the material. The magnetic shields at 4K are made of Amumetal. Since the cryostat is much smaller, there is only one PT420 for the 40K and 4K stage, plus the additional PT420 for the DR.

\section{SIMULATION RESULTS}\label{sec:results}

\subsection{Large Aperture Receiver Telescope}\label{subsec:resLATR}

\begin{figure}[h]
    \centering
    \includegraphics[scale=0.35]{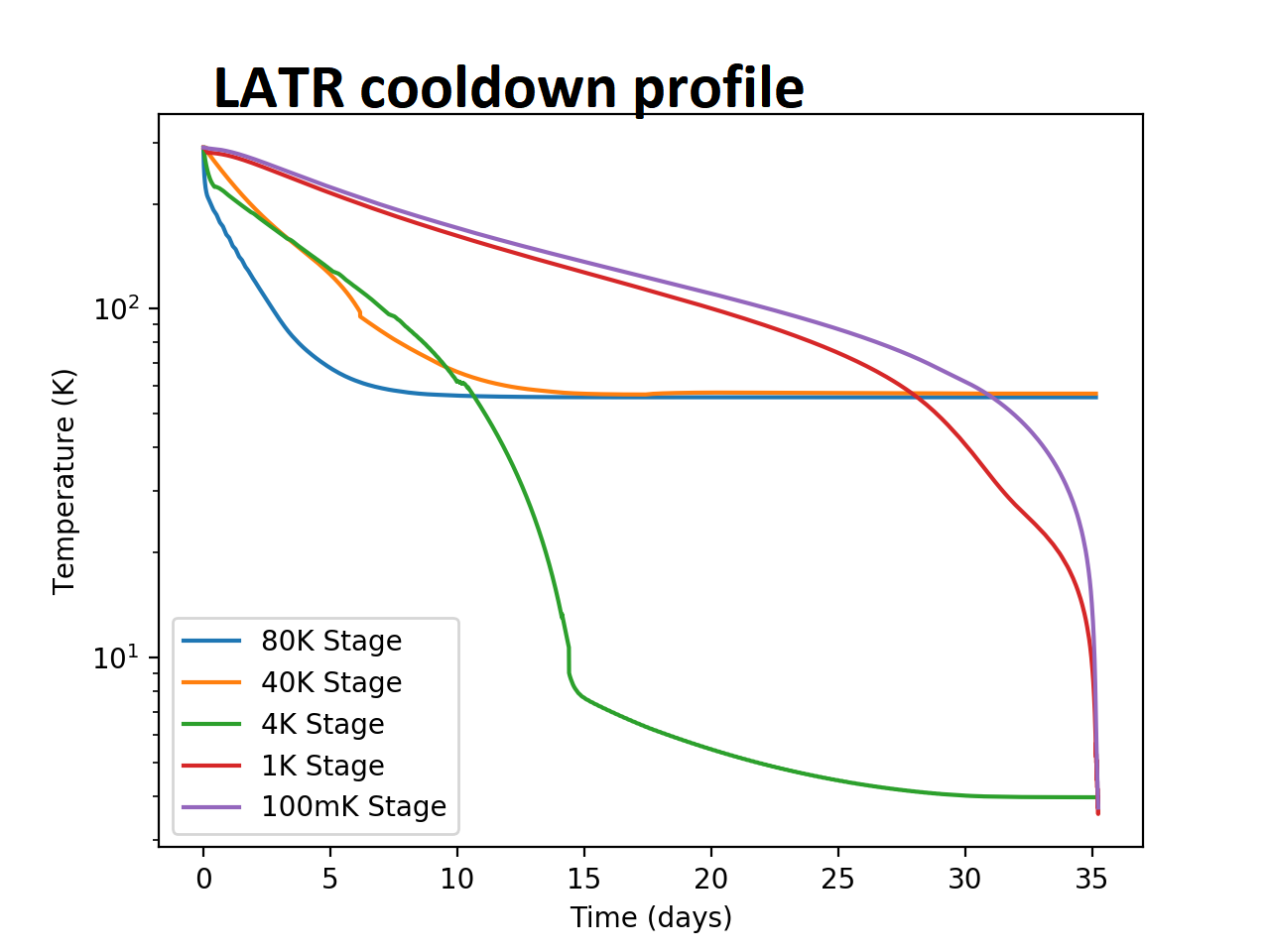}
    \caption{Profile of a single point on the main plate for each stage during the cool down of the LATR cryostat. At approximately 15 days it is possible to notice the change in the gradient of the $4$ K profile. This feature is due by the radiation of the warm $1$ K stage on the $4$ K stage which is not compensated anymore by the cooling power of the PT420 which drops at $10$ K as shown in Fig. \ref{fig:PT420_second}.}
    \label{fig:profile_Large}
\end{figure}

The simulation results of the cooling time for each stage of the LATR to reach equilibrium are presented in table \ref{tab:LATRtime}.
The temperature profiles during the cooldown on the filter plates ($80$ K and $40$ K), main plate ($4$ K) and bus plates ($1$ K and $100$ mK) are presented in Fig. \ref{fig:profile_Large}. From this plot, the influence  of the warm $1$ K and $100$ mK stages on the $4$K stage is evident. Indeed, this stage takes a long time to thermalize due the significant load coming from the $1$ K and $100$ mK stages. It is also possible to note that the $40$ K stage seems hotter than the $80$ K stage. This is due by the fact that the heat strap on the $80$ K stage is connected directly to the plate where the profile is computed, while the heat strap on the $40$ K stage is bolted in the middle of the $40$ K cylinder, which is connected to the plate where the profile is computed.

\begin{table}[h]
	\centering
    \begin{tabular}{lccccc}
    	Stage & $80$ K & $40$ K & $4$ K & $1$ K & $100$ mK \\
        \hline \\
        Time (days) & 10 & 15 & 30 & 35.5 & 35.5 \\ 
    \end{tabular}
\caption{Cool down time for each stage of the LATR.}
\label{tab:LATRtime}
\end{table}

The predicted cool down time of over 35 days for the LATR could have a significant affect on SO scheduling and observation strategies. As a consequence, we are developing different strategies to reduce the cool down time. In particular, we are working on developing heat switches that would harnessing the cooling power available at the warmer stages, such as the 80 K and 40 K stages, at the colder stages. However, different strategies need to be adopted when connecting the $40$ K stage with the $4$ K stage compared to the connection between the $4$ K stage with the $1$ K stage. The first connection will likely be made with nitrogen heat pipes that are being tested, ref. \citenum{Zhu2018}. Whereas, the second connection will probably utilize gas-gap heat switches. The use of gas-gap switches reduces the parasitic load on the colder stages in the off-condition compared to the nitrogen heat pipe.  

\subsection{Small Aperture Telescope}\label{subsec:resSAC}

\begin{figure}[h]
    \centering
    \includegraphics[scale=0.3]{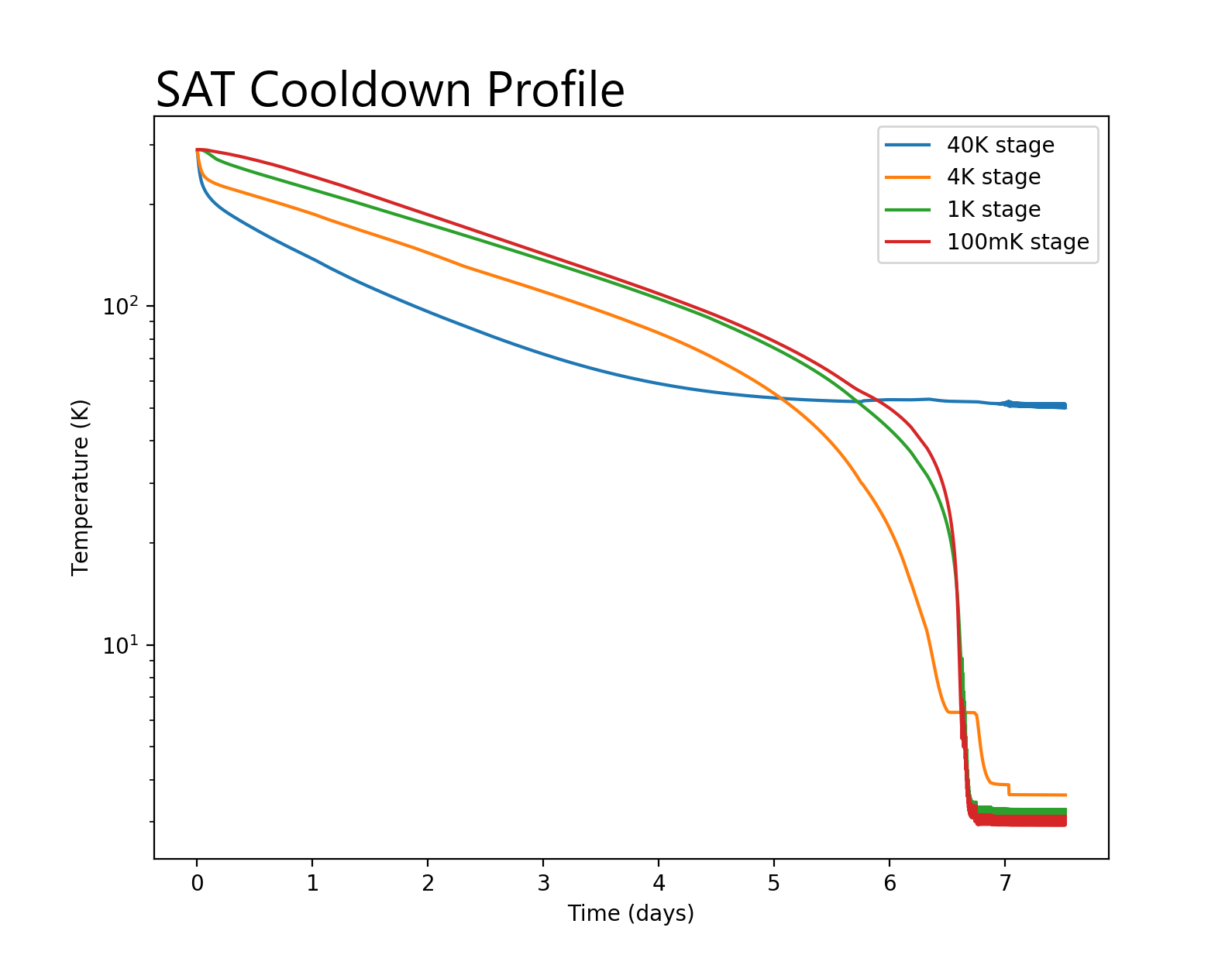}
    \caption{Profile of a single point on the main plate for each stage during the cooldown.}
    \label{fig:profile_small}
\end{figure}

The simulation results of the cooling time for each stage of the SAT to reach equilibrium are presented in table \ref{tab:SATtime}.
For SAT, the rate limiting stage for the cool down is the 4 K stage. This is mainly due to the fact that the majority of the mass is concentrated in this stage. In this setup, the $1$ K and $100$ mK stages are significantly less massive compared to the LATR, therefore the time required to cool them is significantly reduced with respect to the $4$ K stage. 
In figure \ref{fig:profile_small} the cooldown of the cell closest to the heat strap is presented. 
The $40$ K cooldown profile is smoother compared to the one for the LATR because the capacity map for the PT420 was better sampled. The $4$ K stage shows the typical drop in temperature below $50$ K as expected. The coupling between the $1$ K stage and the $4$ K is  evident at $6.5$ days. Indeed, as soon as the $1$ K temperature diminishes the load on the $4$ K is reduced and so the temperature is reduced too.  

\begin{table}[h]
	\centering
    \begin{tabular}{lcccc}
    	Stage & $40$ K & $4$ K & $1$ K & $100$ mK \\
        \hline \\
        Time (days) & 5.5 & 7 & 6.6 & 6.6 \\ 
    \end{tabular}
\caption{Cool down time for each stage of the SAT.}
\label{tab:SATtime}
\end{table}

\section{CONCLUSION}\label{sec:conclusion}

The code developed has allowed us to estimate the cooldown of both the LATR and SAT of the Simons Observatory. The cool down time is particularly important for the LATR.
Indeed, given the dimensions of the LATR, a proper study of the cool down of the cryostat is necessary. The results from this simulation show that the required time to cool down to 4K is approximately 35 days. This time is dominated by cooling the $1$ K and $100$ mK stages. A possible solution to reduce this time is the use of switches connecting different stages to redistribute the cooling power. For example, nitrogen heat pipes for thermally connecting the $40$ K stage with the $4$ K stage are in the testing phase. Whereas, for connecting colder stages, such as the $4$ K stage with the $1$ K stage more classical heat switches, like gas-gap, are considered. Future tests will give a complete characterization of both nitrogen heat pipes and gas-gap switches so that it is possible to include in the cool down simulation. The other stages take less time to thermalize, therefore connecting these colder stages will provide additional cooling power for the $1$ K and $100$ mK stages.
Instead, for the SAT, the cooling time is in the order of $1$ week, which is within SO defined technical requirements. This study could serve as a useful reference for the design of the next generation CMB experiments, like CMB-S4.

\section{Acknowledgments}

This work was supported in part by a grant from the Simons Foundation (Award \#457687, B.K.)

\bibliography{report} 
\bibliographystyle{spiebib} 

\end{document}